\documentclass[journal=jacsat,manuscript=article]{achemso}
\usepackage[version=3]{mhchem}  
\usepackage{bm}
\usepackage{array}    
\usepackage{amsmath}  
\usepackage{longtable} 
\usepackage{url}

\author{Xiongzhi Zeng\textsuperscript{a}}
\affiliation{Hefei National Research Center for Physical Sciences at the Microscale, University of Science and Technology of China, Hefei, 230026,China.}
\author{Huili Zhang\textsuperscript{a}}
\affiliation{Beijing Academy of Quantum Information Sciences, Beijing 100193, China.}
\author{Shizheng Zhang}
\affiliation{Key Laboratory of Precision and Intelligent Chemistry, University of Science and Technology of China, Hefei, 230026, China.}
\author{Pei Liu}
\affiliation{State Key Laboratory of Low Dimensional Quantum Physics, Department of Physics, Tsinghua University, Beijing 100084, China}
\author{Kehuan Linghu}
\affiliation{Beijing Academy of Quantum Information Sciences, Beijing 100193, China.}
\author{Jiangyu Cui}
\affiliation{Department of Modern Physics, University of Science and Technology of China, Hefei 230026, China.}
\author{Xiaoxia Cai}
\affiliation{Beijing Academy of Quantum Information Sciences, Beijing 100193, China.}
\email{caixx@baqis.ac.cn}
\author{Jie Liu}
\affiliation{Hefei National Laboratory, University of Science and Technology of China, Hefei 230088, China.}
\email{liujie86@ustc.edu.cn}
\author{Zhenyu Li}
\affiliation{Key Laboratory of Precision and Intelligent Chemistry, University of Science and Technology of China, Hefei, 230026, China.}
\alsoaffiliation{Hefei National Laboratory, University of Science and Technology of China, Hefei 230088, China.}
\email{zyli@ustc.edu.cn}
\author{Jinlong Yang}
\affiliation{Key Laboratory of Precision and Intelligent Chemistry, University of Science and Technology of China, Hefei, 230026, China.}
\alsoaffiliation{Hefei National Laboratory, University of Science and Technology of China, Hefei 230088, China.}

\title{Accurate Chemical Reaction Modeling on Noisy Intermediate-Scale Quantum Computers Using a Noise-Resilient Wavefunction Ansatz}

\abbreviations{IR,NMR,UV}

\begin{document}
\maketitle
\footnotetext[1]{These authors contributed equally to this work.}

\begin{abstract}
Quantum computing is of great potential for chemical system simulations. In this study, we propose an efficient protocol of quantum computer based simulation of chemical systems which enables accurate chemical reaction modeling on noisy intermediate-scale quantum (NISQ) devices. In this protocol, we combine an correlation energy-based active orbital selection, an effective Hamiltonian from the driven similarity renormalization group (DSRG) method, and a noise-resilient wavefunction ansatz. Such a  combination gives a quantum resource-efficient way to accurately simulate chemical systems. The power of this protocol is demonstrated by numerical results for systems with up to tens of atoms. Modeling of a  Diels-Alder (DA) reaction is also performed on a cloud-based superconducting quantum computer. These results represent an important step forward in realizing quantum utility in the NISQ era. 
\end{abstract}

\section{INTRODUCTION}
In the rapidly evolving field of quantum computing, recent advancements have underscored significant quantum advantages conferred by various devices\cite{arute2019quantum,zhong2020quantum}, positioning quantum chemistry as a prime area for application of this technology.\cite{mcardle2020quantum} Quantum phase estimation algorithm,\cite{kitaev1995quantum} recognized for its potential, encounters practical challenges in real-world applications due to the substantial quantum resources it demands. In the current Noisy Intermediate-Scale Quantum (NISQ) era, the ambition for practical quantum computation faces notable obstacles due to the limitations of available quantum resources, including the restricted number of qubits and notably short quantum coherence times. These limitations, exacerbated by noise, compromise the development of quantum computing capabilities by affecting the reliability and outcomes of quantum simulations.\cite{bharti2022noisy}

Amid these challenges, hybrid quantum-classical algorithms, particularly the variational quantum eigensolver (VQE)\cite{peruzzo2014variational}, emerge as a preferred approach for NISQ computers to address practical problems. Recently, latest quantum hardware and meticulously designed quantum circuits were used to simulate the dissociation curves of molecules such as LiH and F$_2$\cite{guo2022scalable}. However, the use of a minimal basis set limits direct comparison with experimental data. For chemical reactions, the Google team used VQE to study the diazene isomerization reaction on the Sycamore quantum computer at the Hartree-Fock(HF) level of theory without electron correlation\cite{google2020hartree}. Later, homolytic and heterolytic reactions of the H$_3$S$^+$ molecule was studied using a superconducting quantum computer within an active space.\cite{motta2023quantum}. Simulating an active space presents a promising solution to current limitations in quantum resources \cite{ma2023multiscale, von2021quantum}. However, the selection of the active space directly influences the outcomes. Furthermore, the commonly employed quantum circuit ansatz necessitates a substantial number of quantum gates and faces hardware constraints, which complicates achieving high precision.

The effective Hamiltonian approach \cite{motta2020quantum} for NISQ computers focuses on a selected subset of degrees of freedom or energy scales, constructing a reduced Hamiltonian for efficient quantum chemistry simulation. The quantum unitary downfolding formalism, based on the driven similarity renormalization group (DSRG)\cite{huang2023leveraging}, simplifies the treatment of complex quantum systems by reducing the full system Hamiltonian into a lower-dimensional one that retains essential physics, achieving efficient and accurate descriptions of strongly correlated electron systems. Although very efficient, this method has been used to study the bicyclobutane isomerization reaction using only a single qubit, which does not fully reflect the advantages of the method and the characteristics of quantum computers due to the absence of qubit entanglement and quantum circuit design.

In addressing these complexities, this work introduces a novel algorithm for the automatic selection of orbitals based on orbital correlation energy, coupled with an efficient downfolding hamiltonian method that leverages hardware adaptable quantum circuits\cite{zeng2023quantum}.  By amalgamating these two strategies, we have developed a new tiered algorithm capable of conducting high-precision simulations of real systems. Our approach has been validated against established methods for automatic orbital selection, with numerical analysis and testing on model molecular systems confirming its efficacy. Furthermore, incorporating our group’s development of hardware adaptable ansatz has enabled the high-precision simulation of the Diels-Alder (DA) reaction \cite{liepuoniute2024simulation,houk1973generalized} on actual quantum computers, demonstrating the practical feasibility of our proposed scheme. This work represents a significant contribution to the field of quantum computational chemistry by bridging theoretical models and practical applications. It highlights a substantial step forward in utilizing quantum computing for complex chemical reaction simulations, illustrated by the successful simulation of the DA reaction, a pivotal reaction in organic chemistry. The achievements documented herein not only showcase the potential of our methods but also pave the way for future investigations into other complex chemical processes using quantum computing technologies.

The subsequent sections of the paper are organized as follows: In Section II, we primarily outline the workflow of the entire algorithm, which includes the algorithm for automatic orbital selection based on orbital correlation energy, acquiring an effective Hamiltonian from the DSRG method, and finally detailing the hardware adaptable ansatz circuit and computational as well as experimental specifics. In Section III, we commence by comparing various methods for selecting active orbitals, followed by applying the algorithm to the calculation of small molecules and real reaction barriers on classical simulators. Subsequently, we implement this algorithm on actual quantum hardware, achieving promising results. In the concluding section, we summarize the prospective applications of this method.

\section{METHODS}
\begin{figure*}[]
\centering
\includegraphics[width=15cm]{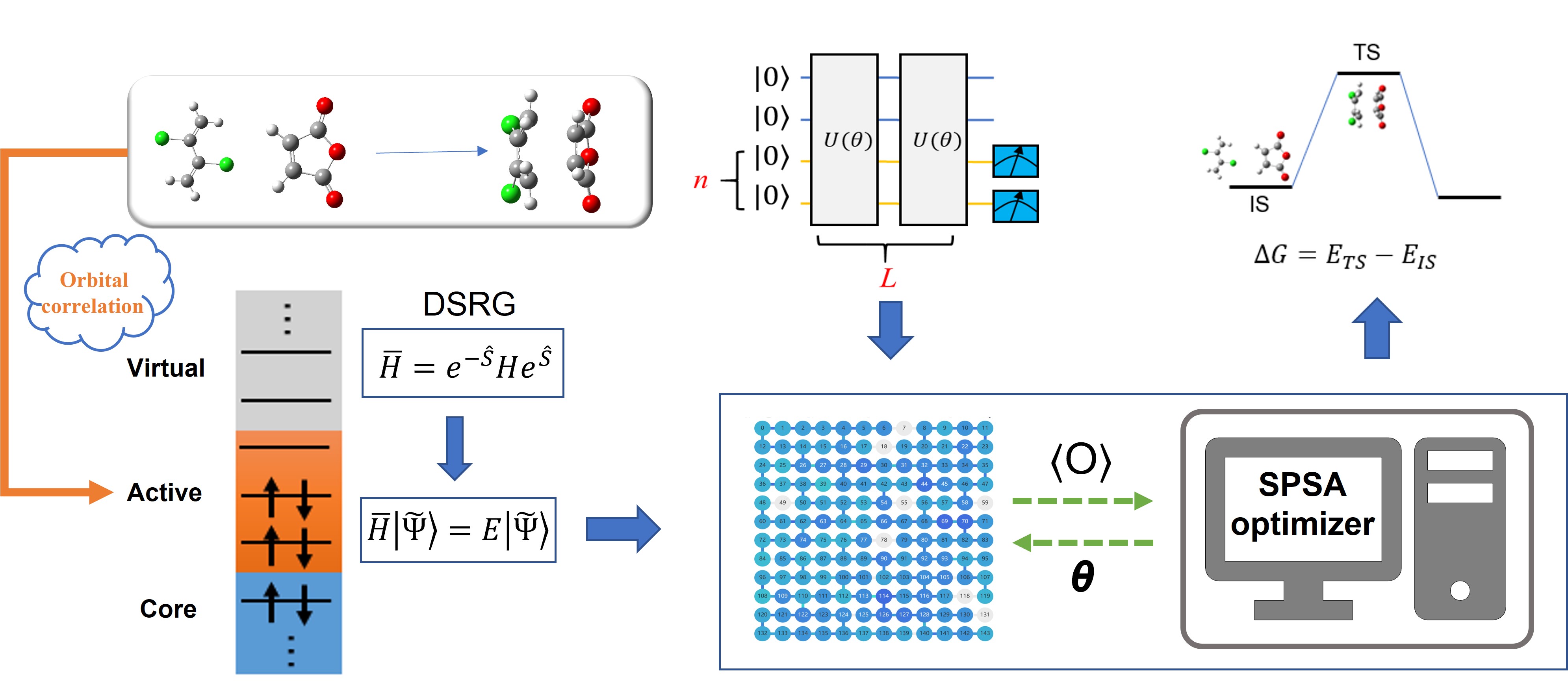}
\caption{Schematic of the workflow for the main algorithms.}
\label{fig1}
\end{figure*}

The algorithm presented in this paper is composed of three main components, as illustrated in \ref{fig1}. Firstly, it employs orbital correlation information to automatically select active orbitals. Following this, the DSRG method is utilized, in conjunction with the selected active orbitals, to construct an effective Hamiltonian. Finally, the efficiency and noise resilience of the hardware adaptable ansatz (HAA) are leveraged to generate the corresponding quantum circuit. By integrating these components, the VQE algorithm can be applied to simulate real reaction systems on either simulators or actual quantum experimental setups. This structured approach ensures a coherent and systematic pathway to achieving high-precision simulations of complex chemical processes, highlighting the synergy between advanced quantum methods and practical quantum computing capabilities.

\subsection{Correlation-Based Orbital Selection Algorithm}
Accurately describing the interactions between electrons is the ultimate goal of post-Hartree-Fock methods. Electrons can occupy different orbitals through excitations, thus multiple configurations in the ground state wave function generally contribute to the energy. By progressively incorporating more orbitals into the active space to calculate energy variations, we can derive the correlation energy between orbitals. This is the main idea behind the many-body expansion full configuration interaction method\cite{eriksen2017virtual,eriksen2019generalized}, where the correlation energy of orbitals can be expressed as follows:
\begin{equation}
	\Delta \epsilon_{i} = \epsilon_{i} - \epsilon_{ref}   
	\label{eq:e1orb}
\end{equation}
\begin{equation}
    \Delta \epsilon_{ij} = \epsilon_{ij} - \Delta \epsilon_{i} -  \Delta \epsilon_{j}  
	\label{eq:e2orb}
\end{equation}
\begin{equation}
	E_{FCI} = \epsilon_{ref} + \sum_{i} \Delta \epsilon_{i} + \sum_{i<j} \Delta \epsilon_{ij} +... 
	\label{eq:e1orb}
\end{equation}
Where the $\epsilon_{ref}$ is the reference energy which can be from the HF or complete active space configuration interaction (CASCI) calculation. The $\Delta \epsilon_{i}$ and $\Delta \epsilon_{ij}$ are the correlation energy from one and two orbitals respectively. Numerical calculations demonstrate that this method achieves high precision in both weakly correlated and strongly correlated systems\cite{eriksen2018many,eriksen2019many}, making it an important method for calculating ground state energy. However, the complexity of the calculations grows exponentially with the increase in the number of electrons n and orbitals N. Another point of interest is that the contribution from correlation energy tends to decrease sharply with the order of expansion, with the most significant contributions often coming from the initial few orders.\cite{eriksen2020ground}  

Inspired by the autoCAS method\cite{stein2019autocas,stein2016automated}, which uses the orbital entanglement entropy calculated from an approximate ground state wave function as a criterion for selecting active orbitals, we propose using the single and double orbital correlation energy $\Delta \epsilon_{i}$ and $\Delta \epsilon_{ij}$ obtained from the many-body expanded full configuration interaction method as a basis for selection. We determine the size of the active space based on the relative contributions of orbital energies. An orbital with a significant individual energy contribution is considered active, and if substantial correlation energy exists between two orbitals, both are deemed active. In our method, the highest occupied molecular orbital (HOMO) and the lowest unoccupied molecular orbital (LUMO) are automatically included because they are directly related to the molecule's reactivity. During the orbital selection process, we first sort all orbitals by the absolute values of their energies from highest to lowest, then select those with larger contributions, and calculate their relative contributions to the correlation energy. The only adjustable parameter is the size of this relative contribution, and various simulations have shown that 30$\%$ is an appropriate value. For specific molecules, adjustments can be made based on the required quantum hardware resources, ensuring that important orbitals are included.
The calculation cost is polynomial $O(N^2)$ and highly parallel and these correlation energies are directly associated with the final correlation energy instead of an approximation. Our numerical simulation results show (see below) that the active orbitals obtained with this method are highly consistent with those obtained through autoCAS and projection-based methods, while requiring no prior knowledge, having a low computational demand, being highly parallelizable, and the selected orbitals being closely related to the electron correlation energy.

\subsection{Downfolded Hamiltonian}
Effective Hamiltonian theory has been posited as an innovative method to lessen the qubit requirements\cite{yanai2006canonical,white2002numerical}. This theory focuses on "downfolding" the system Hamiltonian, simplifying the complex many-body problem into a manageable form that still captures the essence of the quantum system's behavior. One of the representatives is DSRG method\cite{evangelista2014driven,li2019multireference}, which the Hamiltonian can be written as: 
\begin{equation}
	\hat{H}(s) = \hat{U}^{+}(s)H\hat{U}(s)=e^{-\hat{A}(s)}He^{\hat{A}(s)}   
	\label{eq:eham}
\end{equation}
where the $\hat{H}(s)$ is the transformed Hamiltonian and $\hat{U}(s)$ is a unitary with a continuous parameter s which defined in the range [0, $\infty$). $\hat{A}(s)$ is written in terms of the standard cluster operator $\hat{T}(s)$ as $\hat{A}(s) = \hat{T}(s) -\hat{T}^{+}(s)$, where $\hat{T}(s)$ is a sum of excited operators $\hat{T}_{k}(s)$, with k ranging from one to the total number of electrons(n)
\begin{equation}
	\hat{T}(s) = \hat{T}_{1}(s)+\hat{T}_{2}(s)+... \hat{T}_{n}(s)  
	\label{eq:Top}
\end{equation}
In the DSRG, the similarity Hamiltonian is driven by the source operator $\hat{R}(s)$, according to the following a equation:
\begin{equation}
        \left[ \hat{H}(s) \right]_{od} =\left[ e^{-\hat{A}(s)}He^{\hat{A}(s)}  \right]_{od} = \hat{R}(s) 
	\label{eq:dsrg}
\end{equation}
The equation is augmented with an appropriate boundary conditions for the operator. For s = 0, the non-diagonal component of the Hamiltonian is identical to the bare Hamiltonian $\left[ \hat{H}(0)  \right]_{od} = H_{od}$ a condition that can be trivially satisfied if S(0) = 0. For s $\rightarrow{\infty}$, the DSRG flows decouple excited configurations from the reference, driving the off-diagonal part of $\hat{H}(s)$ to zero, $\left[ \hat{H}(\infty)  \right]_{od} = 0$. The DSRG method necessitates the delineation of a decaying source operator and a flow parameter, facilitating the iterative derivation of the transformed Hamiltonian to subsequently ascertain the energy. This technique preliminarily consolidates the correlation effects of electrons exterior to the active space into an transformed Hamiltonian, the eigenvalues of which yield the system's energy. The precision of the ground state energy computed via the DSRG method is situated between that of CCSD and CCSD(T)\cite{li2015multireference}.

\subsection{Ansatz Design and Optimization}
Once the Hamiltonian is determined, another critical challenge is how to design an appropriate wave function ansatz or quantum circuit\cite{cao2019quantum}. Although the DSRG method can reduce the number of qubits required for simulating systems, short coherence times and noise remain pressing issues that must be addressed on current NISQ hardware\cite{mcardle2020quantum}. While reducing circuit depth can partly solve the problem of short coherence times, commonly used wave function ansatz, such as hardware efficient ansatz(HEA)\cite{kandala2017hardware}, unitary coupled cluster ansatz\cite{anand2022quantum}, the adaptive derivative-assembled pseudo-trotter method\cite{grimsley2019adaptive} and the qubit coupled-cluster method\cite{ryabinkin2018qubit}, lack noise-resistant features. Our group has developed a hardware adaptable ansatz \cite{zeng2023quantum}, inspired and improved upon from quantum neural networks, with its corresponding quantum channel defined as
\begin{equation}
        \rho^{out} = tr_{anc}(U(\rho^{in}\bigotimes \rho^{anc})U^{+})
	\label{eq:haa}
\end{equation}
\begin{equation}
        U=\prod^{L}_{l=1}\prod^{N}_{i=1} U^{l}_{i}(\theta^{l}_{i})
	\label{eq:haaU}
\end{equation}
where $\rho^{in}, \rho^{anc}, \rho^{out}$ are the density matrix of the system, ancilla and output respectively and $\rho^{in}/\rho^{anc}=|00..0><0..00|$. The $tr_{anc}$ means partial trace over the ancilla layer. N is the number of ancilla qubits and L is the number of entangled layers. We have discovered that the precision of convergence and the ability to resist noise can be enhanced by increasing the number of auxiliary qubits; the actual number of auxiliary qubits required depends on the real hardware. Our numerical tests indicate that the introduction of these ancillary qubits significantly enhances the circuit's resistance to noise\cite{zeng2023quantum}. Moreover, the HAA has been identified as an efficient, low-depth circuit model adept at reproducing the ground-state energy of quantum systems. Its adaptability and resource efficiency make it a compelling model for quantum computations in the context of limited qubit availability.

\subsection{Simulation Details and Experimental setup}
We have employed the quantum algorithm simulation platform, Q$^2$Chemistry\cite{fan2022q}, developed by our group to implement the entire algorithmic workflow. The source code of the algorithm and the key results of the paper are publicly available on GitHub\cite{zeng001}. For most classical computations, such as HF, MP2, NEVPT2, CASSCF, and FCI, we utilize the Pyscf software\cite{sun2018pyscf}. The DSRG method and the extraction of its effective Hamiltonian are performed using the Forte software\cite{evangelista2014driven,li2019multireference}. In mapping fermions to qubits, we have employed the Bravyi-Kitaev transformation\cite{seeley2012bravyi} and reduced the requirement by two qubits leveraging symmetry. For classical optimization, we use the Adam algorithm\cite{kingma2014adam} to optimize the parameters of the quantum circuit. For VQE simulation, a layer-wise optimization strategy is also adopted. For the quantum noise simulation for HAA and HEA, the depolarizing error model was used.

We perform the VQE experiment on on superconducting qubits selected from BAQIS’s quantum cloud Baiwang (SCQ-P136)\cite{quafu_2024,quafu_2023,quafu001} with simultaneous perturbation stochastic approximation (SPSA) optimization \cite{Kandala_2017} on a classical computer. The circuit utilized for minimizing the ground state energy of initial and transition states are designed by HAA, which includes 19 adjustable rotation gates along $Z$ axis and $X$ axis with 17 parameters and 4 CNOT gates (see below). After iteration, we apply  zero-noise extrapolation (ZNE)\cite{Kim_2023_1,Kim_2023_2,Foss_2022} approach to suppress experimental errors. In our ZNE approach, each single CNOT gate in the circuit is replaced by $N$ copies of itself, while the single qubit gates are in consistent with the gates in corresponding trial steps. The final energy is obtained by linear extrapolating the averaged energy of 10 steps to the noiseless limit $N =0$.

\section{RESULTS}
\subsection{Comparative Analysis of Automated Orbital Selection Algorithms}
We begin to introduce an automatic active space selection method based on orbital correlation energy, and compared with other methods based on projection and orbital entropy by calculating the dissociation energy and potential energy surface (PES) of the nitrogen molecule.The calculation of nitrogen molecule's ground state energy and PES stands as a quintessential benchmark for assessing quantum chemistry methods, given its simple diatomic composition juxtaposed with a complex electronic structure marked by a triple bond. This dual simplicity and complexity render it an exemplary model for gauging the precision and efficiency of varied quantum chemistry approaches. Furthermore, this calculation also emerges as a pivotal test case for evaluating quantum algorithms and hardware performance, offering insights into quantum computing's prowess and constraints in tackling intricate chemical quandaries.

\begin{figure*}[]
\centering
\includegraphics[width=12cm]{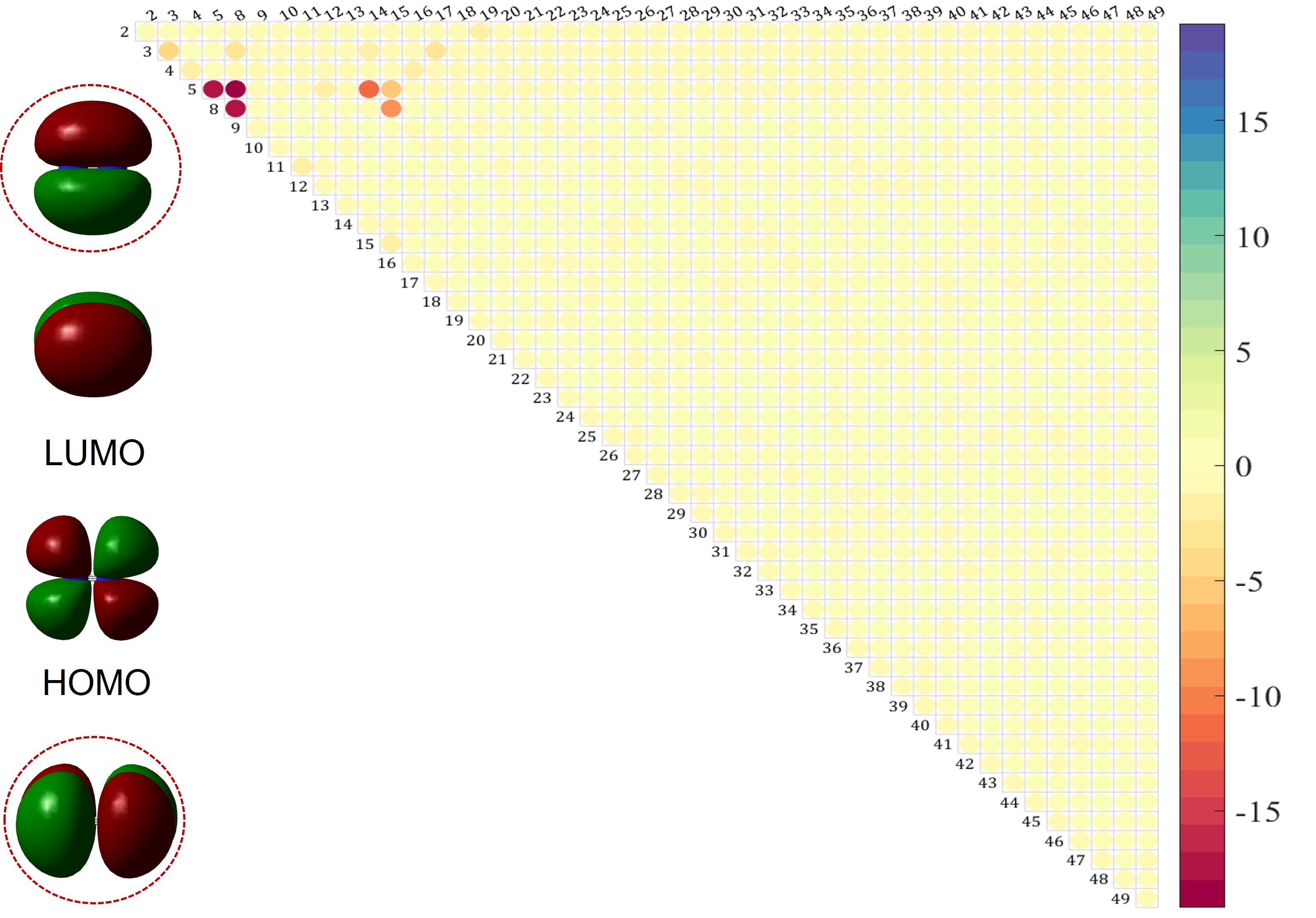}
\caption{The one and two orbital correlation energy the nitrogen molecule with bond length 1.09 $\AA$.}
\label{fig_n2}
\end{figure*}

We've computed single and double orbital correlation energies for $N_2$ on equilibrated bond lengths, as depicted in Fig \ref{fig_n2}. This elucidates the varying correlation degrees among distinct orbitals, predominantly marked by a lack of correlation—a sparse matrix representation that underpins our active orbital selection. Notably, orbitals proximal to the HOMO and LUMO manifest substantial correlation energies, intricately linked to the reaction's bond formation and dissociation dynamics. In proximity to the equilibrium position, a pronounced hybridization of bonding and antibonding orbitals is observed, contrasting with the predominantly discrete atomic orbitals at the dissociation limit.

We compared the results obtained from our method with two currently successful approaches: the projection-based AVAS and the orbital-entropy-based AutoCAS, as illustrated in Table \ref{bs2}. Our approach yields active orbitals closely matching those obtained by the other methods, and especially relative to the autoCAS method. The AutoCAS method is based on orbital entropy which is very similar to orbital correlation energy and can produce the most relevant active orbitals. However, its performance rely heavily on the precision of the DMRG wave function. On the other hand, the AVAS method, requiring only the valence orbitals of atoms, can significantly increase the active space for systems with extensive orbital hybridization. Our orbital-energy-based selection method does not necessitate extensive prior knowledge or parameters, enabling automated operation and large-scale parallel computation, directly correlating the selected orbitals with the necessary correlation energy calculations, thus facilitating subsequent computations. 

\begin{table}[H]
    \caption{The active orbital index and dissociation energy for nitrogen molecule with from CASCI, CASSCF and CASSCF+NEVPT2 and DSRG at bond length of 1.09 $\AA$ and 6.0 $\AA$. The unit for energy is eV}
    \begin{longtable}{cccccccc}
     \hline
     Methods & 1.09 $\AA$ & 6.0 $\AA$ & CASSCI  & CASSCF  & NEVPT2 & DSRG  & Exp.   \\
     \hline
        AVAS      & [3,4,5,6,7,8,9]  & [4,5,6,7,8,9]   & 8.3262  & 9.0100   & 9.4416 &  9.748 &  \\
        autoCAS   & [5,6,7,8]        & [4,5,6,7,8,9]   & 8.0977  & 7.6335   & 9.6511 &  9.747 & 9.765  \\
        This work & [5,6,7,8]        & [4,5,6,7,8,9]   & 8.0977  & 7.6335   & 9.6511 &  9.747 &  \\
    \hline       
    \end{longtable}
    \label{bs2}
\end{table}

From Table \ref{bs2}, it is evident that different selections of orbitals yield distinct interaction energies, with the main differences arising from ignored interactions with other orbitals. These overlooked interactions can be accounted for through orbital optimization, such as with the CASSCF method. Table \ref{bs2} reveals that while orbital optimization does lower the system's energy, its impact remains considerably constrained. An alternative method involves applying perturbative corrections, for example, the NEVPT2 approach. As shown in Tables \ref{bs2}, perturbative additions are able to bring the dissociation energy of nitrogen in line with experimentally comparable magnitudes, in contrast to orbital optimization, which causes the dissociation energy to deviate from values comparable to experimental data\cite{frost1956dissociation}. Simultaneously,we note that the DSRG method, which preemptively folds the remaining orbitals into the active space, can also produce results that closely match experimental data, further substantiating the viability of our method. Finally, We have calculate the PES with the three methods as in Figure S2 and the results indicate that the energy outcomes derived from the three methods are in close agreement, which further corroborates the precision and utility of our proposed method.



\subsection{Classical simulation for Diels-Alder reaction}
\begin{figure*}[]
\centering
\includegraphics[width=15cm]{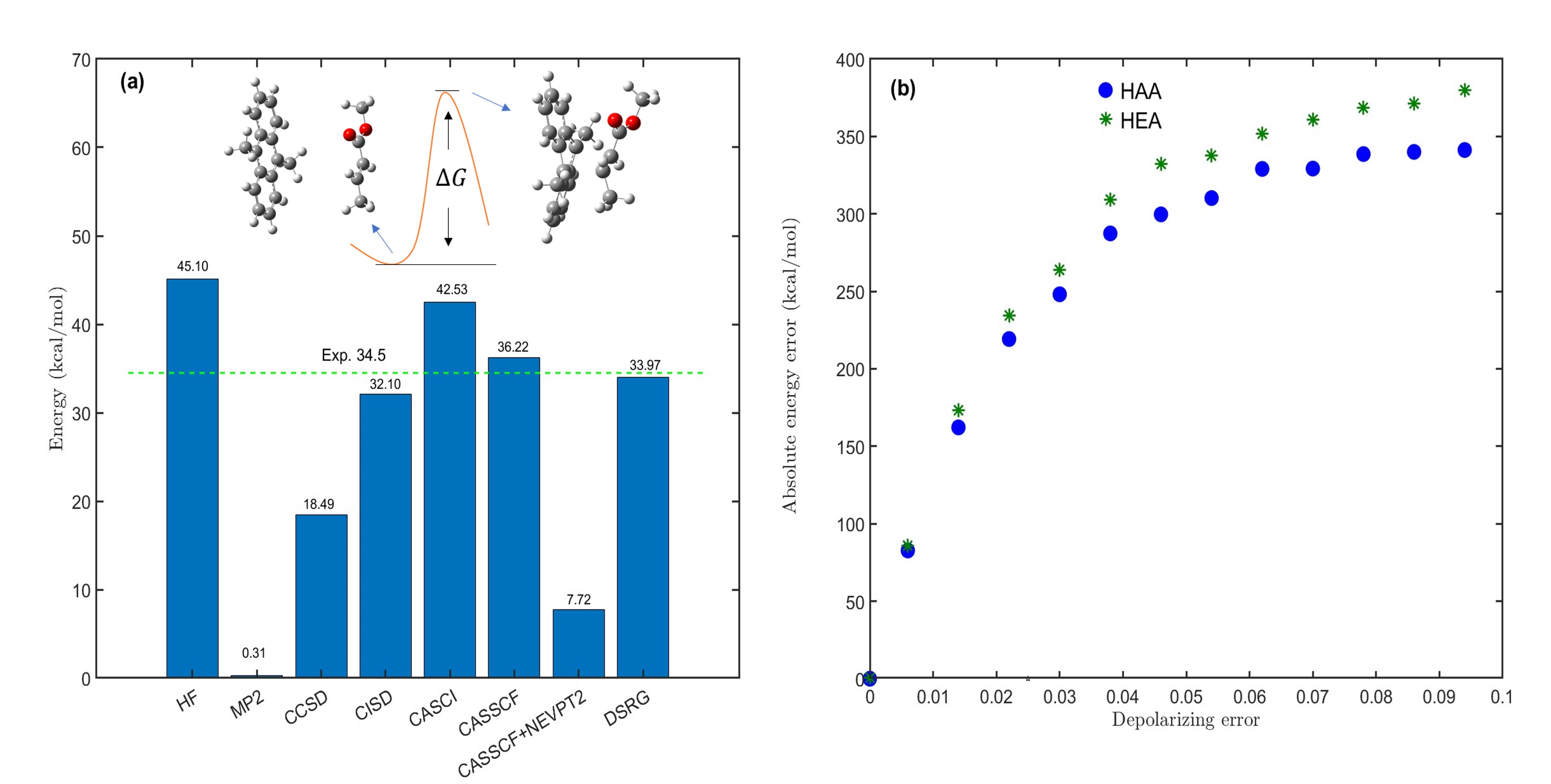}
\caption{(a)The reaction barrier calculated by HF, MP2, CCSD, CASCI, CASSCF, CASSCF+NEVPT2 and DSRG. (b) The classical quantum noise simulation with HAA and HEA ansatz for initial state.}
\label{fig_da_1}
\end{figure*}

In order to explore the practicality of our current algorithm and assess its potential application in handling more complex chemical systems, we extended its application to the calculation of reaction barriers involving larger systems. Specifically, we selected a complex DA reaction system consisting of 45 atoms, involving the molecules C$_{5}$H$_{8}$O$_{2}$ and C$_{16}$H$_{14}$ as inset in \ref{fig_da_1}, containing 164 electrons and 366 orbitals with the 6-31g(d) basis set. Through in-depth analysis, we obtained the active space with 4 orbital and 4 electron based on the orbitals' correlation energies, and these orbitals demonstrate significant hybridization and correlation among the orbitals, providing a solid testing ground for our algorithm.

We initially employed a variety of common classical wave function methods as in \ref{fig_da_1}(a), including single-reference state-based HF, CCSD, and multiconfigurational approaches like configuration interaction with single and double excitations (CISD), CASCI, and CASSCF. Additionally, we incorporated some perturbative methods, including second-order Møller-Plesset (MP2) and NEVPT2. We found that for this reaction, multiconfigurational methods, such as CASSCF, could achieve relatively precise results, for instance, a reaction energy of 36.22 kcal/mol, which is very close to the experimental value of 34.5 kcal/mol\cite{tang2012accurate}. Although adding perturbative methods introduced significant deviations, using DSRG method yielded highly accurate results, with an energy of 33.97 kcal/mol.

Notably, with a refined selection of orbitals, our algorithm required consideration of only 4 key orbitals. Utilizing the effective Hamiltonian calculated through the DSRG method, combined with 2 auxiliary qubits, a quantum circuit with 8 qubits with 24 two-qubit gates was sufficient to achieve chemical accuracy in line with experimental data as in SI. To test the intrinsic noise-resistance capabilities of the HAA, we conducted a comparison with an HEA circuit that has the same number of single and double qubit gates as the current HAA. We found that, without the aid of any other error suppression techniques, the HAA circuit was able to reduce errors through the use of auxiliary qubits. These results demonstrate the potential of the HAA for real quantum computers. These results not only validate the efficiency and accuracy of our algorithm in processing larger systems but also demonstrates its immense potential in the field of precise chemical simulation.

\subsection{Quantum experiment verification for our algorithm}
\begin{figure*}[]
\centering
\includegraphics[width=13cm]{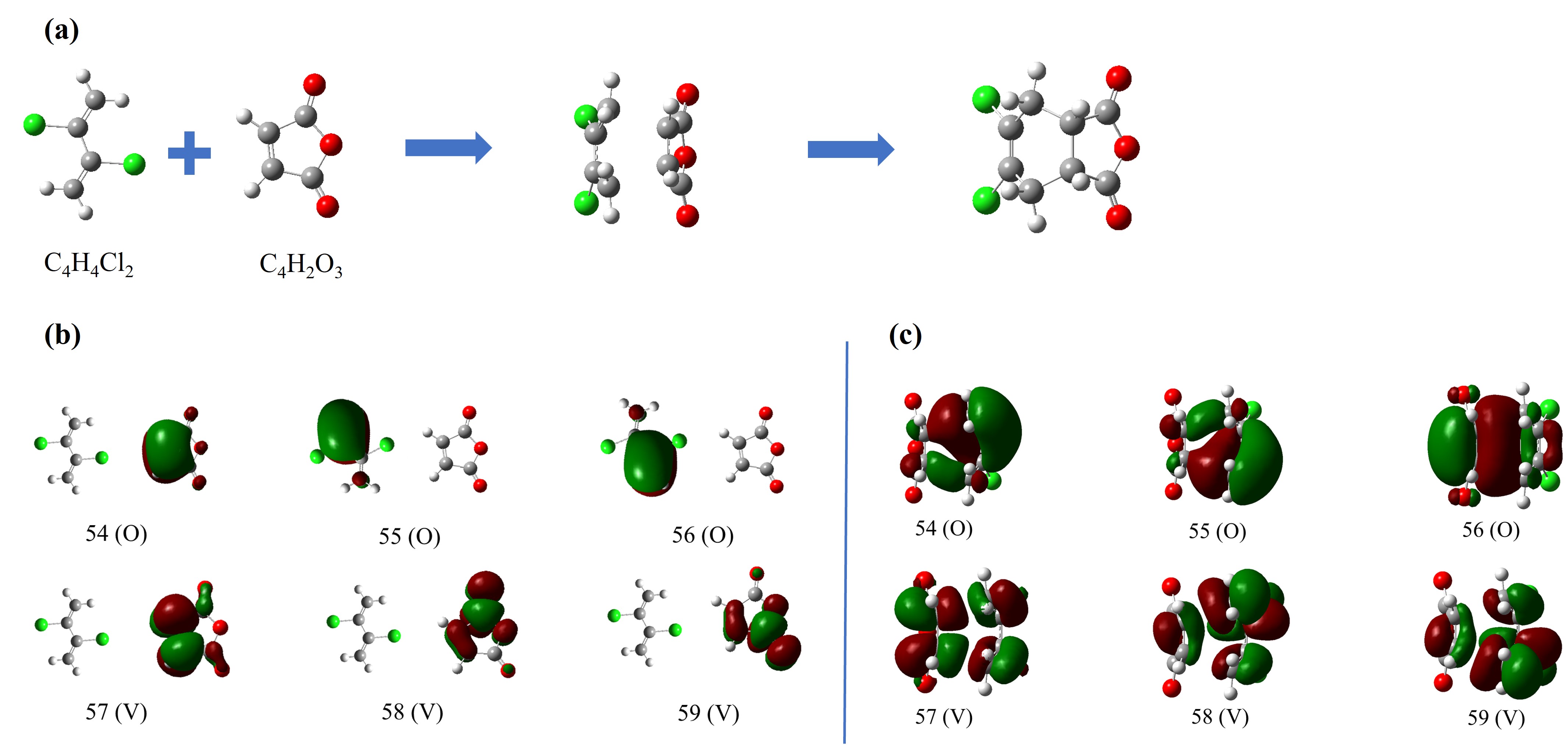}
\caption{The Diels-Alder reaction between C$_{4}$H$_{4}$Cl$_{2}$ and C$_{4}$H$_{2}$O$_{3}$ molecules (a) geometry structure for the whole reaction. The selected orbitals for (b) initial state and (c) transition state for the reactions.}
\label{fig_geo_da}
\end{figure*}

Our automated orbital selection algorithm offers an approach for the practicality of quantum computing. We have use a cloud-based quantum computer to simulate another Diels-Alder reaction between C$_{4}$H$_{4}$Cl$_{2}$ and C$_{4}$H$_{2}$O$_{3}$ (maleic anhydride) molecules as in \ref{fig_geo_da}, a key reaction in current pharmaceutical synthesis. We take accurate prediction of its activation energy barrier as our experimental goal. 

As demonstrated in Figure \ref{fig_geo_da} (b) and (c), we have analyzed the orbitals for initial and transition states of this system, which includes 112 electrons and necessitates dealing with 220 orbitals under the cc-pvdz basis set. We observed significant orbital correlations, especially apparent in the transition state with pronounced hybrid orbital features. The highest correlation levels were noted in the HOMO and LUMO orbitals, leading us to select the two most active orbitals for further computations, and employing the DSRG method, we obtained a folded Hamiltonian.

\begin{table}[]
    \caption{The calculated and experimental energy for Diels-Alder reaction with maleic anhydride with different methods including CASSCF, DSRG.}
    \begin{longtable}{ccccc}
     \hline
      Energy (Ha) & CAS(6o,6e) & CAS(2o,2e) & DSRG                   \\
     \hline
       IS         &-1450.052175   & -1450.025308 & -1451.899418        \\
       TS         &-1450.038176   & -1449.975190 & -1451.855420       \\
       Barriers (Kcal/mol)  &8.78 & 31.45        & 27.61                \\
    \hline
        Exp. (Kcal/mol)    && 26.50    \\
     \hline
    \end{longtable}
    \label{exp_data}
\end{table}

\begin{figure*}[htbp]
\centering
\includegraphics[width=16cm]{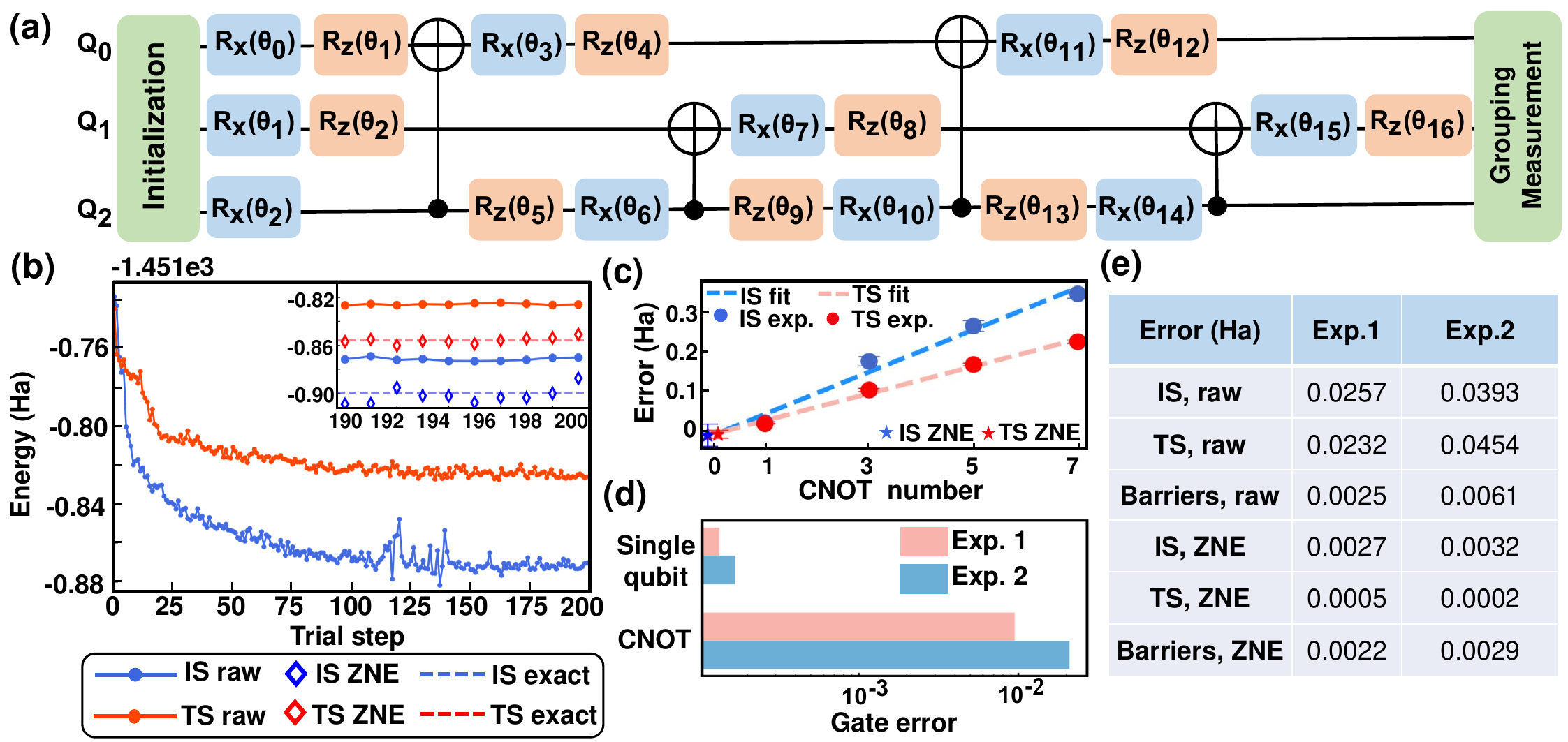}
\caption{(a) The encoded circuit of hardware adaptable ansatz. The single qubit rotation angles $\theta_{i},i\in[0,16]$ are parameters to be optimized in the experiment. (b) The energy minimization procedure for IS energy (blue) and TS energy (red). The inset on the top shows the raw energy and energy after ZNE for the final 10 steps. (c) The linear fitting of the averaged IS energy error and TS energy error of the final 10 steps, with each CNOT gate replaced by 1,3,5, and 7 CNOT gates respectively. The correlation coefficients for the linear fittings are 0.9738 (IS) and 0.9916 (TS). (d) Average single qubit gate errors and CNOT gate errors of two qubit groups in experiments. (e) The energy error comparison of HAA experiments with two qubit groups. }
\label{fig_circuit}
\end{figure*}

For the quantum circuit, a simplified HAA circuit is used as shown in Figure \ref{fig_circuit}a, incorporating four two-qubit gates and an auxiliary qubit. Experiments revealed the reaction barrier to be merely 26.5 kcal/mol, presenting a significant challenge for both classical and quantum simulations that necessitate precise modeling. Classical simulation outcomes indicated that improper active space selection, like CASSCF(6,6), could result in substantial error (18 kcal/mol) in estimating the reaction barrier, impeding direct comparisons with experimental data\cite{tang2012accurate}. However, our method of orbital selection based on correlation energy achieved commendable accuracy, with a mere 5 kcal/mol deviation, further affirming the utility of our orbital selection approach. By integrating the DSRG method with quantum simulations, we managed to further minimize the error to within 1 kcal/mol.

The experiment was performed on a cloud-based hardware, \textit{Quafu Baiwang}. The energy optimization process is shown in Figure \ref{fig_circuit}b. To mitigate experimental noises, the ZNE approach is applied, and the result is shown in Figure \ref{fig_circuit}c. After ZNE, the error of the energy can be suppressed below $3\times10^{-3}$, closing to the threshold of chemistry accuracy. To validate our method's feasibility further, we perform the experiment with another group of qubits on the same chip. The error rates of average single qubit gates and CNOT gates in two experiments are shown in Figure \ref{fig_circuit}d. The comparisons of energy errors in two experiments shown in Figure \ref{fig_circuit}e demonstrate that our scheme is accurate even when the average CNOT gate error rate exceeds 1\%, further attesting to HAA's noise resistance against noisy quantum circuits.  Moreover, compared with HEA, introducing an extra qubit in HAA still yield results comparable to those without auxiliary qubits, underscoring HAA's compatibility with imperfect quantum operations. The convergence process, albeit challenging due to the multitude of controlling factors in current quantum computers, excitedly suggests an energy deviation of a few mHa, particularly for the VQE convergence results of the transition state, maintaining stable absolute energy errors. The variance in final reaction barriers further corroborate the entire algorithm's practicality. With the addition of common noise-reduction algorithms like ZNE, our findings could be further refined.

\section{Conclusion}
To fully leverage the current NISQ computers for processing real chemical reactions, our team has developed a unique algorithm. The core innovation of this algorithm lies in its clever integration of several key technologies: Firstly, by selecting active orbitals based on orbital correlation energy, we ensure that only the most impactful orbitals are involved in the quantum computing process, thereby reducing computational complexity. Next, using the driven similarity renormalization group method to derive an effective Hamiltonian, this theoretical foundation enables us to accurately capture the complex interactions among electrons in chemical systems. Additionally, we introduce a hardware adaptable wavefunction ansatz, further enhancing the algorithm's compatibility and efficiency on actual quantum hardware. We have illustrated the application of this algorithm in processing larger-scale and more complex chemical reaction systems through classical simulators, and further validating it on real quantum computing hardware. we demonstrated its feasibility, showcasing its efficiency and accuracy in handling specific chemical reactions. These results indicate that our algorithm is not only suitable for NISQ quantum computers but also has the potential to become an important tool in chemical simulation and materials science research as quantum computing technology matures. In summary, our algorithm opens a new avenue for simulating real chemical reactions using NISQ quantum computers, not only improving computational efficiency and accuracy but also laying a solid foundation for the future application of quantum computing in the fields of chemistry and materials science. As quantum computing technology continues to advance, we anticipate that this algorithm will play a key role in solving more complex chemical problems, propelling scientific research in related fields into a new stage.

\begin{acknowledgement}
This work is supported by the Strategic Priority Research Program of the Chinese Academy of Sciences(XDB0450101), the National Natural Science Foundation of China (22303090, 22393913, 22303005, 92365206), Innovation Program for Quantum Science and Technology (Grant No. 2021ZD0303306, 2023ZD0300200, 2021ZD0301802) and the USTC Supercomputing Center.

\end{acknowledgement}

\begin{suppinfo}

\end{suppinfo}

\bibliography{dsrg+haa}

\end{document}